\documentclass[prb,reprint,twocolumn,superscriptaddress,noshowpacs,notitlepage,longbibliography,10pt,citeautoscript]{revtex4-2}%

\usepackage{pifont}
\usepackage{pgffor}
\usepackage{graphicx,bm,times}
\usepackage{amsmath}
\usepackage{amsfonts}
\usepackage{amssymb}
\usepackage{mathtools}
\usepackage{color}
\usepackage{hyperref}
\hypersetup{
	colorlinks = true,
	allcolors = {blue}
}

\newcommand{\up}{\uparrow}
\newcommand{\dn}{\downarrow}

\begin{document}

\title{Persistence of charge ordering instability to Coulomb engineering\\
in the excitonic insulator candidate TiSe\textsubscript{2}}

\author{Sebastian Buchberger}
\affiliation{SUPA, School of Physics and Astronomy, University of St Andrews, St Andrews KY16 9SS, United Kingdom}
\affiliation{Max Planck Institute for Chemical Physics of Solids, N\"othnitzer Strasse 40, D-01187 Dresden Germany}

\author{Yann in 't Veld}
\affiliation{I. Institute of Theoretical Physics, Universit{\"a}t Hamburg, D-22607 Hamburg, Germany}

\author{Akhil Rajan}
\affiliation{SUPA, School of Physics and Astronomy, University of St Andrews, St Andrews KY16 9SS, United Kingdom}

\author{Philip~A.~E.~Murgatroyd}
\affiliation{SUPA, School of Physics and Astronomy, University of St Andrews, St Andrews KY16 9SS, United Kingdom}

\author{Brendan Edwards}
\affiliation{SUPA, School of Physics and Astronomy, University of St Andrews, St Andrews KY16 9SS, United Kingdom}

\author{Bruno~K.~Saika}
\affiliation{SUPA, School of Physics and Astronomy, University of St Andrews, St Andrews KY16 9SS, United Kingdom}

\author{Naina Kushwaha}
\affiliation{SUPA, School of Physics and Astronomy, University of St Andrews, St Andrews KY16 9SS, United Kingdom}
\affiliation{STFC Central Laser Facility, Research Complex at Harwell, Harwell Campus, Didcot OX11 0QX, United Kingdom}

\author{Maria H. Visscher}
\affiliation{SUPA, School of Physics and Astronomy, University of St Andrews, St Andrews KY16 9SS, United Kingdom}
\affiliation{Max Planck Institute for Chemical Physics of Solids, N\"othnitzer Strasse 40, D-01187 Dresden Germany}

\author{Jan Berges}
\affiliation{U Bremen Excellence Chair, Bremen Center for Computational Materials Science, and MAPEX Center for Materials and Processes, University of Bremen, D-28359 Bremen, Germany}

\author{Dina Carbone}
\affiliation{MAX IV Laboratory, Lund University, Lund, Sweden}

\author{Jacek Osiecki}
\affiliation{MAX IV Laboratory, Lund University, Lund, Sweden}

\author{Craig Polley}
\affiliation{MAX IV Laboratory, Lund University, Lund, Sweden}

\author{Tim Wehling}
\email[Correspondence to: ]{tim.wehling@physik.uni-hamburg.de}
\affiliation{I. Institute of Theoretical Physics, Universit{\"a}t Hamburg, D-22607 Hamburg, Germany}
\affiliation{The Hamburg Centre for Ultrafast Imaging, D-22761 Hamburg, Germany}

\author{Phil~D.~C.~King}
\email[Correspondence to: ]{pdk6@st-andrews.ac.uk}
\affiliation{SUPA, School of Physics and Astronomy, University of St Andrews, St Andrews KY16 9SS, United Kingdom}

\date{\today}

\begin{abstract} 
TiSe\textsubscript{2} has long been considered one of the best candidate materials to host the elusive excitonic insulator (EI) phase. However, a finite coupling to the lattice can generically be expected, while a lack of ``smoking-gun'' signatures for the importance of the electron-hole interaction in driving the phase transition has rendered it challenging to distinguish the EI from the conventional charge-density-wave (CDW) phase. Here, we demonstrate a new approach, exploiting the susceptibility of excitons to dielectric screening. We combine mechanical exfoliation with molecular-beam epitaxy to fabricate ultra-clean van der Waals heterostructures of monolayer \mbox{(ML-)TiSe$_2$/graphite} and ML-TiSe$_2$/hBN. We observe how the modified substrate screening environment drives a renormalisation of the quasi-particle band gap of the TiSe$_2$ layer, signifying its susceptibility to Coloumb engineering. The temperature-dependent evolution of its electronic structure, however, remains unaffected, indicating that excitons are not required to drive the CDW transition in TiSe\textsubscript{2}.

\end{abstract}

\maketitle

\section{Introduction}
\begin{figure}[t!]
    \centering
    \includegraphics[width=0.85\columnwidth]{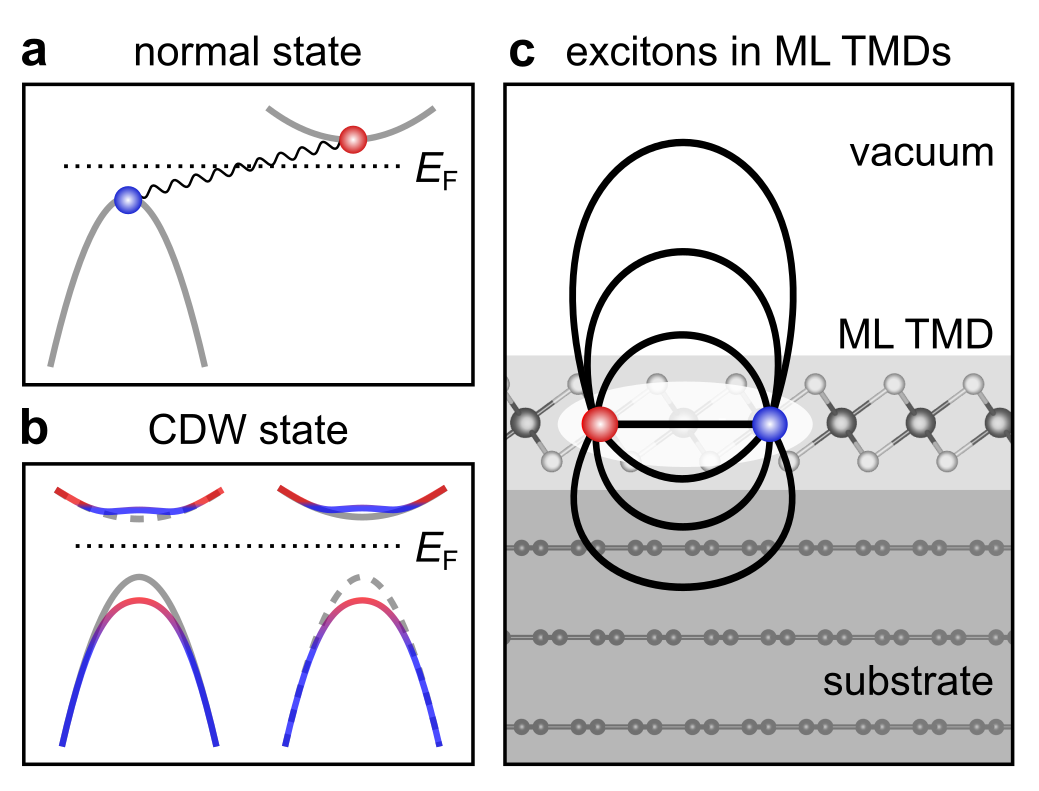}
    \caption{{Electronic structure and ordering instability in TiSe$_2$.} (a,b) Simplified schematic low-energy electronic structure in the (a) normal state and (b) ordered state of TiSe$_2$. The pairing interactions that could mediate this transition in an excitonic insulator model are shown schematically in (a). In (b), solid (dashed) grey lines represent the original (replica) version of the bands in the normal state. The coloured lines represent the electronic structure in the ordered state, with blue (red) colour indicating the band character of the normal state electron (hole) band. (c) Illustration of reduced screening of the electron-hole interactions in the monolayer limit. The degree of change expected will be modified both by dielectric screening from the encapsulating media and from internal free-carrier screening within the 2D layer itself.}
    \label{fig:fig1}
\end{figure}

For the past half decade, TiSe\textsubscript{2} has been one of the most actively-studied and controversial members of the family of transition metal dichalcogenides. Its normal state electronic structure is best described as an indirect-gap semiconductor with close to zero band gap (Fig.~\ref{fig:fig1}(a))~\cite{Cercellier2007,Monney2010,Watson2019,Larsen2024}. This is, in principle, the ideal band structure to support the formation of an excitonic insulator (EI), an exotic state where electrons and holes spontaneously form excitons and Bose condense, driving a transition into a charge ordered state~\cite{Jerome1967,Kohn1967}. The observation of a charge-density wave (CDW) instability occurring at around 200~K therefore generated significant excitement, and TiSe$_2$ has been proposed as one of the most likely candidate materials to host the EI phase \cite{DiSalvo1976,Wilson1977,Pillo2000,Cercellier2007,Monney2010}. However, this mechanism has been questioned in favour of a simple structural phase transition \cite{Hughes1977, Rossnagel2002, Wegner2020}, where the charge density merely follows the periodic lattice distortion due to electron-phonon coupling. Importantly, both mechanisms break the same underlying symmetries of the crystal, and give rise to essentially identical electronic structures in the ordered state (Fig.~\ref{fig:fig1}(b)); it is therefore extremely challenging from studying either the crystal or electronic structure to establish which of excitonic or lattice contributions plays the dominant role in driving the phase transition. Despite numerous attempts to answer this question -- for example, by probing the collective excitations of TiSe$_2$~\cite{Kogar2017,Lin2022} or the response of its ordering instability to ultrafast optical excitation~\cite{Rohwer2011,Porer2014,Monney2016,Kurtz2024} -- the origin of the phase transition in TiSe$_2$ remains highly controversial.

\begin{figure*}
    \centering
    \includegraphics[width=0.85\textwidth]{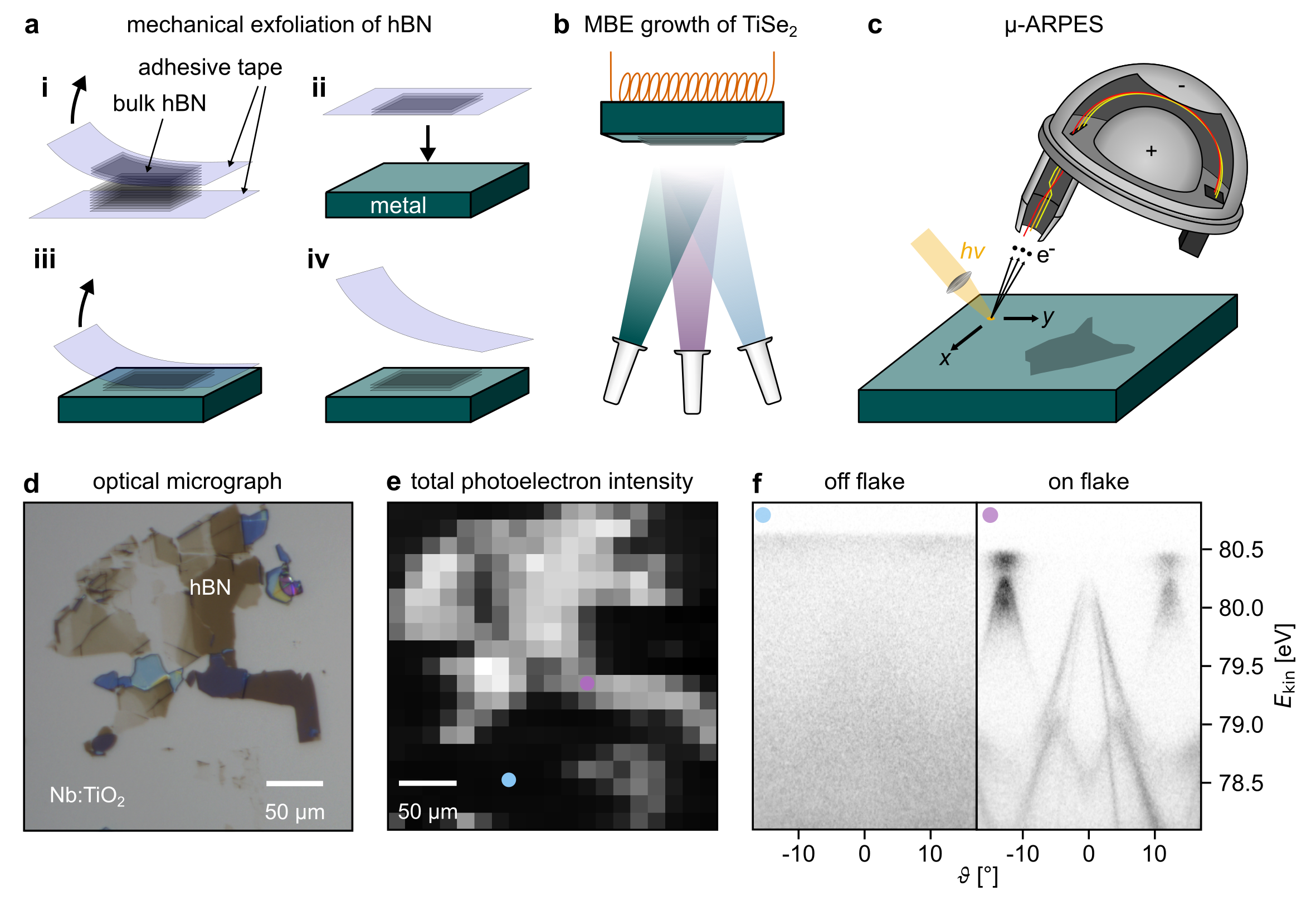}
    \caption{{ML-TiSe\textsubscript{2}/hBN heterostructures.} (a--c) Fabrication and characterisation: (a) hBN bulk crystals are exfoliated to flakes with a thickness of several 10's of nm, which are then transferred on to a conducting substrate; (b) Monolayer films of TiSe\textsubscript{2} are grown atop the hBN flakes by molecular beam epitaxy; (c) Electronic structure measurements of the TiSe\textsubscript{2}/hBN heterostructures are enabled by $\mu$-ARPES. (d) Optical micrograph of hBN flakes on a conducting Nb:TiO\textsubscript{2} substrate. (e) Spatial map of integrated photoemission intensity in the region around the large flake shown in (d). (f) Representative photoemission measurements performed with the beam located off (left) and on (right) the hBN flake. High-quality ARPES spectra of ML-TiSe\textsubscript{2} are only obtained from atop the hBN surface.
    }
    \label{fig:fig2}
\end{figure*}

Here, we develop a new approach: we study monolayers of TiSe$_2$, where we statically modify the dielectric screening by control of the environment. It is well known from studies of, e.g., semiconducting WS$_2$ and MoSe$_2$ that electron-hole interactions can be dramatically enhanced by reduced screening in the 2D limit (see Fig.~\ref{fig:fig1}\,(c))~\cite{Ye2014, Chernikov2014,Ugeda2014}, yielding a tuning parameter that is unique to 2D materials. To date, however, it remains unclear whether this provides an effective tuning strategy for the CDW phase in TiSe\textsubscript{2} \cite{Goli2012,Li2016a,Chen2016,Kolekar2018,Kolekar2018a}, with studies typically finding a very similar transition temperature in monolayer (ML) samples as for the bulk~\cite{Chen2015,Kolekar2018a,Watson2020,Sugawara2016,antonelli2023controlling}.

\section{Hybrid exfoliation-epitaxy of ML-T\MakeLowercase{i}S\MakeLowercase{e}\textsubscript{2}-based van der Waals heterostructures}
In order to realise an optimal low-screening environment for ML-TiSe\textsubscript{2}, it should be fabricated on an insulating support substrate, with the ``gold-standard'' material being hexagonal boron nitride (hBN). However, this presents two significant experimental challenges: 
(1)~ML-TiSe$_2$ has never been realised via mechanical exfoliation, with exfoliated samples in the literature typically having thicknesses on the order of 5--10~nm~\cite{Li2016}. Instead, ML-TiSe$_2$ samples are typically fabricated via molecular-beam epitaxy~\cite{Chen2015, Sugawara2016, Watson2020, Rajan2020}. Obtaining high-quality epitaxial transition-metal dichalcogenide (TMD) layers on hBN substrates is notoriously challenging due to the slugish growth kinetics and lack of unsaturated dangling bonds on the substrate surface, typically leading to undesirable growth morphologies and premature onset of bilayer formation~\cite{Seredynski2022}. 
(2)~Arguably the most effective route to investigate the phase transition in the monolayer limit of TiSe$_2$ is via the corresponding evolution of its electronic structure, as probed by angle-resolved photoemission (ARPES). However, ARPES cannot typically be performed on insulators like hBN.

\begin{figure*}
    \centering
    \includegraphics[width=0.85\textwidth]{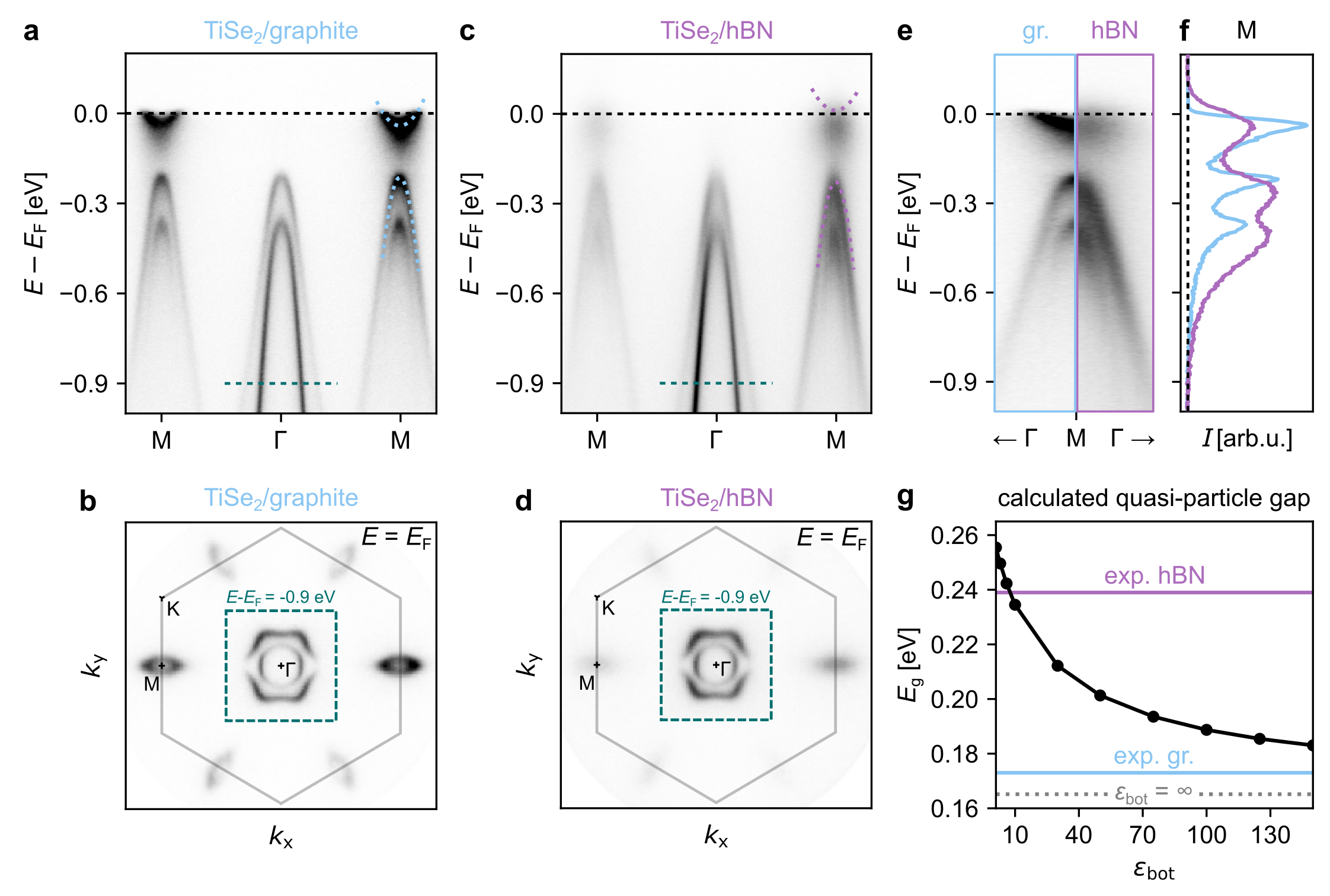}
    \caption{{Low Temperature electronic structure measurements of TiSe\textsubscript{2}/hBN and TiSe\textsubscript{2}/graphite heterostructures.} (a) Band dispersion and (b) constant energy contours at the Fermi level for monolayer TiSe\textsubscript{2}/graphite, measured at $T=18$\,K. The dashed region in (b) shows a constant energy contour at a binding energy of 0.9~eV, showing the valence band (see dashed horizontal line in (a)). (c,d) Equivalent spectra measured on TiSe\textsubscript{2}/hBN. (e) Direct comparison of conduction and valence band spectra at the M point, and (f) corresponding EDCs at the M point. (g) Calculated dependence of the quasi-particle band gap of ML-TiSe\textsubscript{2} on the dielectric environment. $\varepsilon_\mathrm{bot}$ is the dielectric constant of the semi-infinite space below the TiSe\textsubscript{2} layer. The purple and blue lines indicate the quasi-particle gaps for ML-TiSe\textsubscript{2} on hBN and graphite substrates as determined from our experimental ARPES data, and the grey line is the calculated limit for $\varepsilon_\mathrm{bot} = \infty$.
    }
    \label{fig:fig3}
\end{figure*}

To overcome both of these challenges, we have developed a hybrid bottom-up and top-down fabrication procedure, as outlined in Fig.~\ref{fig:fig2} (see Appendix for details). Thin flakes (several 10's of nm thickness) of hBN are exfoliated using the Scotch-tape method and transferred onto a conducting 0.5\,wt\% Nb-doped TiO\textsubscript{2} substrate (Fig.\ref{fig:fig2}\,(a)). Such flakes are thin enough that the photo-induced current across the flake is sufficient to allow ARPES measurements with only a small influence of charging~\cite{Henck2017, Koch2018}.
Next, a monolayer of TiSe\textsubscript{2} is grown on the substrate by molecular beam epitaxy (MBE, Fig.~\ref{fig:fig2}\,(b))). To overcome the challenges of hBN as a growth substate, we use our recently developed method for nucleation-assisted MBE growth of 2D materials~\cite{Rajan2024}, which allows us to fabricate a uniform monolayer with near complete coverage and negligible bilayer regions (see Supplementary Fig.~1 for additional materials characterisation). Due to the small lateral size of the exfoliated hBN flakes (typically on the scale of a few 10\,$\mu$m), electronic structure measurements of the resulting ML-TiSe\textsubscript{2}/hBN heterostructures necessitate microfocus ($\mu$)-ARPES experiments with a beam spot focused down to $\sim$10\,$\mu$m (Fig.~\ref{fig:fig2}\,(c), see Appendix). This provides sufficient spatial resolution to probe the ML-TiSe$_2$/hBN heterostructure, while still allowing good spectral resolution for studying the electronic states of interest: as shown in Fig.~\ref{fig:fig2}\,(d)-(e), the hBN flakes can be readily located from the integrated ARPES contrast in spatial mapping data, due to the distinct photoemission intensity between spectra measured from atop the hBN flake and on the TiO\textsubscript{2} substrate. Comparing the $\mu$-ARPES spectra off and on the flake (Fig.\ref{fig:fig2}\,(f)), it is clear that high quality band dispersions of ML-TiSe\textsubscript{2} are obtained from atop, and only atop, the hBN. These are almost free from charging effects, as evident from the fact that the Fermi levels are nearly aligned in the two measurements shown. Off the flake, only diffuse photoemission signal is obtained, indicating that a well ordered TiSe\textsubscript{2} film is only present on the hBN flakes.

\section{Influence of growth substrate on the ground state electronic structure}
Our hybrid fabrication approach opens the door to a systematic investigation of the effect of the substrate screening on the electronic structure of TiSe$_2$ by comparing to measurements of more conventional TiSe\textsubscript{2}/graphite reference samples. We show measurements of the ground state electronic structure of these systems in Fig.~\ref{fig:fig3}. In both cases, the well-known dispersion of ML-TiSe\textsubscript{2} is obtained, with a pair of spin-orbit-split Se~4$p$-derived valence bands at the $\mathrm{\Gamma}$ point, Ti\,3$d$-derived electron pockets located in the vicinity of the Fermi level at the M points, and strong replicas of the valence band below the Ti~conduction bands. The replica bands, in particular, are a characteristic spectroscopic signature of the CDW state, and arise due to a backfolding of the valence bands by the $2\times2$ periodic lattice distortion that accompanies the CDW (see Ref.~\cite{Watson2020} for a detailed discussion of the ground-state electronic structure). 

While the band features are qualitatively consistent for the samples grown on different substrates, several differences are evident. The spectra from the sample on graphite exhibit nearly symmetric intensity in $k_{x}$, while we observe a clear asymmetry between the backfolded states at the M points located at positive and negative momentum for the sample on hBN. We assign this to the suppression of twin domains in the growth of the TiSe\textsubscript{2} film on hBN, due to the three-fold symmetry of the hBN surface as compared to the sixfold symmetry of a graphene layer.  More importantly, while the spectra in the graphite case show strong, well defined elliptical conduction pockets with a conduction band minimum located approximately $40$\,meV below the Fermi level, we only observe faint tails of the conduction band minimum in the case of the hBN substrate. From two-dimensional fits of the spectral function (see Supplementary Fig.~2), we estimate that the conduction band minimum is located $(12\pm15)$\,meV above the Fermi level for ML-TiSe$_2$/hBN.

Notably, in contrast to previous measurements of monolayer TiSe\textsubscript{2}, our constant energy contours in Fig.~\ref{fig:fig3}\,(b,d) are free from rotational disorder. For the sample on a graphite substrate, the well-developed Fermi contours allow making a precise quantification of the free carrier density of $(3.0\,\pm0.1)\times10^{13}$\,$e^-$/cm$^2$ in the TiSe\textsubscript{2} layer from a Luttinger analysis. The carrier density is clearly significantly lower for the sample grown on hBN. Given the difficulty of precisely locating its essentially unoccupied conduction band, we estimate the carrier density by assuming the Fermi level sits 3~meV above the bottom of the band - the most occupied value allowed by the error bars of our 2D fitting analysis. This yields a generous upper limit of the carrier density for our ML-TiSe$_2$/hBN of $3\times10^{12}$\,$e^-$/cm$^2$, i.e., at least an order of magnitude lower than for the sample on graphite.

While it was previously suggested that off-stoichiometries may be responsible for the electron doping of monolayer TiSe\textsubscript{2} films~\cite{Sugawara2016}, the low concentrations of Se vacancies found in sister samples to those studied here~\cite{Rajan2024} cannot account for the doping level present in the TiSe\textsubscript{2}/graphite sample. Furthermore, the samples on both growth substrates were grown under the same conditions (substrate temperature, metal, and chalcogen flux), and so could be expected to have equivalent vacancy concentrations. We therefore attribute the increased carrier doping in TiSe$_2$/graphite as an influence of intrinsic charge transfer from the semimetallic graphite substrate to the TiSe$_2$ layer atop.

Crucially, this leads to a situation here where {\it both} the dielectric screening from the substrate as well as the internal screening from free carriers in the TiSe$_2$ layer itself is reduced for the TiSe$_2$/hBN heterostructure as compared to the TiSe$_2$/graphite heterostructure. Interestingly, we find that the quasiparticle band gap is larger for TiSe$_2$/hBN than for TiSe$_2$/graphite. If the change in band filling led simply to a rigid shift of the chemical potential, all states would move to lower binding energy. Fig.~\ref{fig:fig3}(e,f), however, shows that the valence bands are shifted to higher binding energy for TiSe$_2$/hBN, despite the smaller filling of the conduction band. From our fits (Supplementary Fig.~2) we find that the band gap in TiSe$_2$/hBN is $(239\pm{15})$~meV, while that for TiSe$_2$/graphite is $(173\pm{6})$~meV, nearly 30\,\% smaller.

In doped semiconductors, it is well established that the increased screening from induced free carriers can lead to a reduction in the quasiparticle band gap with increasing carrier density~\cite{Berggren1981,Bennett1990}, an effect which can be particularly pronounced in the semiconducting TMDs~\cite{steinhoff2014influence,Chernikov2015,Gao2017,erben2018excitation-induced,Qiu2019,sahoo2025quasiparticle}. To confirm whether substrate screening could also drive the band gap changes observed here, we estimate the quasi-particle band gap $E_g$ for different substrate dielectric constants $\varepsilon_{\text{bot}}$, using a model based on a Wannierized density functional theory (DFT) bandstructure \cite{schobert2024ab} with corrections from the Coulomb hole plus screened exchange (COHSX) approximation (see Appendix). As shown in Fig.~\ref{fig:fig3}(g), we find a significant enhancement of $E_g$ as the environmental screening is reduced from a metallic ($\varepsilon_{\text{bot}} = \infty$) to an insulating substrate (lower $\varepsilon_{\text{bot}}$, see also Supplementary Fig.~S3). Using a realistic value of $\varepsilon_{\text{bot}} = 3$ to represent the hBN substrate and $\varepsilon_{\text{bot}} = \infty$ to represent the graphite substrate, we find band gaps of $250$~meV and $165$~meV, respectively. These are comparable to the respective experimentally observed gaps, such that we attribute the change of band gap predominantly to an environmental screening-induced modification of the Coulomb-driven band gap renormalization. Together, our theoretical and experimental findings thus indicate that our samples grown on different substrates are susceptible to ``Coulomb engineering''~\cite{rosner2016two-dimensional,raja2017coulomb,vanLoon2023} of their electronic states.

\begin{figure*}
    \centering
    \includegraphics[width=0.9\textwidth]{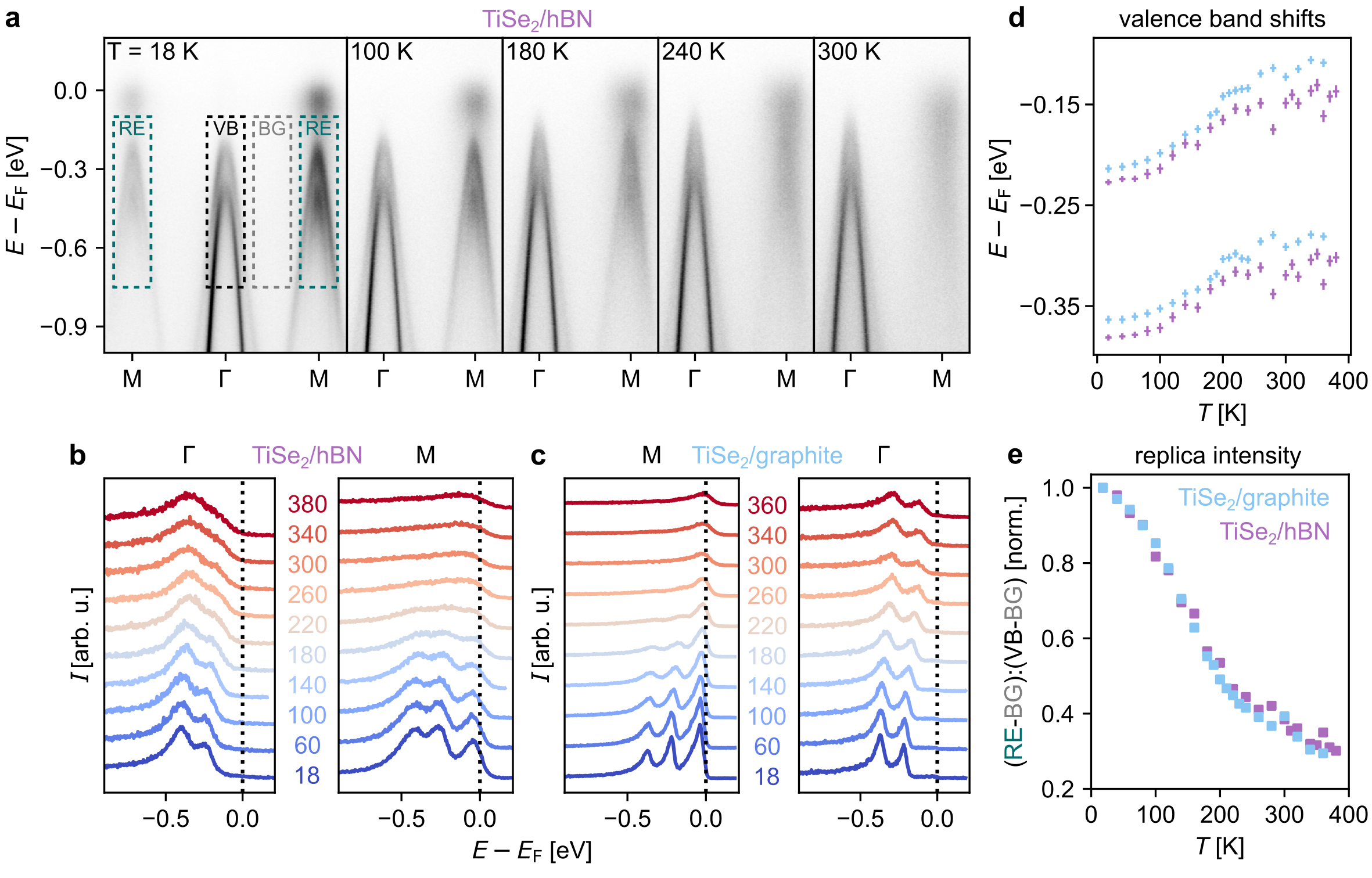}
    \caption{{Temperature-dependent electronic structure of TiSe\textsubscript{2}/hBN as compared to TiSe\textsubscript{2}/graphite.} (a) Band dispersions of TiSe$_2$/hBN taken at selected temperatures across the CDW transition. (b,c) EDCs at selected selected temperatures for (b) TiSe\textsubscript{2}/hBN and (c) TiSe\textsubscript{2}/graphite, showing suppression of replica peaks and valence band shifts. (d) Temperature-dependent evolution of the valence band positions, showing the onset of valence band hybridisation at the CDW transition and (e) evolution of the CDW replica intensity. Both show temperature-dependent trends which show negligible dependence on the substrate.
    }
    \label{fig:fig4}
\end{figure*}

\section{Temperature-dependent evolution of electronic structure}
To investigate whether this has any influence on the phase transition in TiSe$_2$, we have performed temperature-dependent measurements of the electronic structure. Fig.~\ref{fig:fig4}(a) shows the measured dispersions from TiSe\textsubscript{2}/hBN (see Supplementary Fig.~4 for the equivalent measurements of TiSe\textsubscript{2}/graphite). With increasing temperature, we observe a monotonic reduction of the spectral weight of the replica states at the M points, indicating the suppression of the CDW order at higher temperatures. This is also apparent from the energy distribution curves (EDCs) extracted at the M-point, while EDCs taken at the $\Gamma$-point indicate a small temperature-dependent shift of the top of the valence band to higher binding energies with decreasing temperature (Fig.~\ref{fig:fig4}\,(b)). Intriguingly, while the replica states lose sharpness at ~200\,K, significant spectral weight persists to room temperature and beyond. However, precursor fluctuations of the order parameter can persist well above $T_\mathrm{c}$ for a strong-coupling transition, and the features observed from TiSe$_2$/hBN are all qualitatively similar to those observed from our companion samples grown on graphite substrates (Fig.~\ref{fig:fig4}(c), Supplementary Fig.~4) as well as previous measurements for samples on graphene substrates (Refs.~\cite{Watson2020, Chen2015}).

To assess whether there are any signatures of differences in the ordering instability for TiSe$_2$/hBN and TiSe$_2$/graphite, we therefore perform two separate quantitative analyses of the temperature-dependent ARPES data. First, we fit the valence bands positions extracted from temperature-dependent EDCs at $\mathrm{\Gamma}$ (Fig.~\ref{fig:fig4}\,(d)), searching for the onset of shifts due to band hybridisation at the CDW transition~\cite{Watson2020}. While the band positions for TiSe$_2$/hBN are offset to slightly higher binding energies than for TiSe$_2$/graphite, reflecting the larger quasiparticle gap of the former as discussed above, both systems present a nearly indistinguishable temperature evolution. Crucially, both show a change of slope at $\sim$200\,K, suggesting that both share a very similar CDW transition temperature to that of the bulk. Second, we quantify the evolution of the replica spectral weight via a region of interest (ROI) analysis (see dashed lines in Fig.~\ref{fig:fig4}\,(a) and Appendix for details). From Fig.~\ref{fig:fig4}(e), it is clear that both TiSe$_2$/hBN and TiSe$_2$/graphite exhibit an essentially identical CDW behaviour. For both, we observe significant replica weight persisting to our highest measurement temperature of 380\,K, indicative of strong CDW fluctuations. A subtle change of slope at $T\approx\!200$\,K again indicates a phase transition close to the bulk TiSe$_2$ value of 202\,K \cite{DiSalvo1976} in both ML-TiSe$_2$/hBN and ML-TiSe$_2$/graphite. In fact, the persistence of replica weight above the CDW transition is arguably even more evident in the case of graphite (see Fig.~\ref{fig:fig4}\,(c)), where weak valence band peaks are visible in the EDCs at M up to ~300\,K.

\section{Discussion and Perspectives}
Our observation of a large band gap renormalisation of the TiSe$_2$ states of $>60$~meV indicates a significant influence of dielectric screening by the graphite substrate and by the free carriers within the TiSe$_2$ layer itself. Indeed, the free carrier density in the  TiSe$_2$ layer for our TiSe$_2$/graphite 
heterostructures is comparable to the exciton Mott density observed in well-known large-gap semiconducting TMDs even in low screening environments~\cite{Mak2013,Chernikov2015,Steinhoff2017,Lin2019}. In combination with the extrinsic screening from the graphite substrate~\cite{Steinhoff2017}, we thus conclude that our ML-TiSe$_2$/graphite sample is well above the Mott transition and excitons therefore cannot play any significant role in driving a CDW phase transition in this sample. Yet, its temperature-dependent electronic structure evolution is virtually indistinguishable from that of TiSe$_2$/hBN, while the extracted transition temperatures are within experimental error of one another, and of the bulk. Our measurements thus directly indicate that excitons are not required to obtain the well-known CDW phenomenology of TiSe$_2$.

\begin{figure}
\includegraphics[width=.9\linewidth]{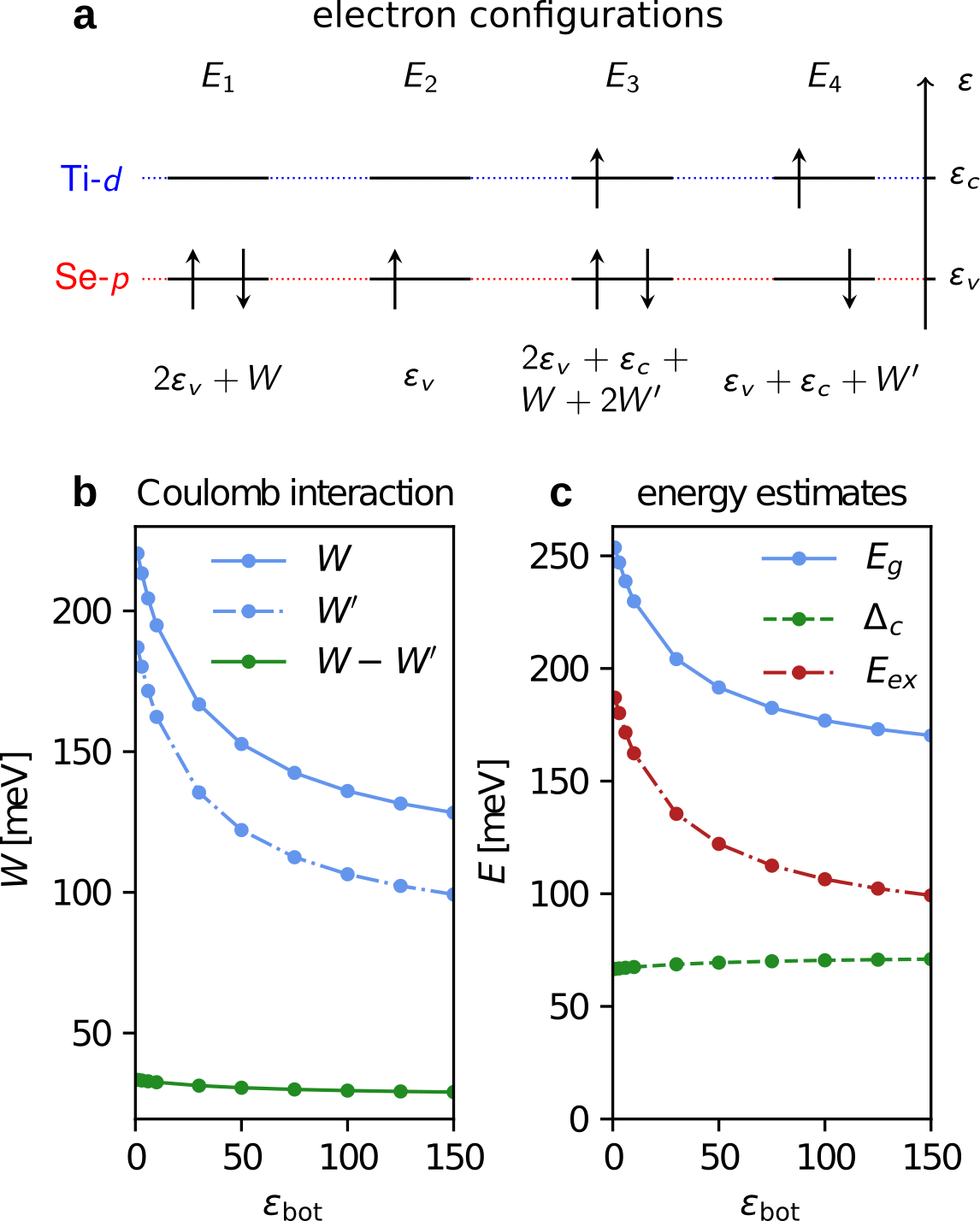}
\caption{{Theoretical modelling of exciton binding energies.} (a) Schematic of four relevant electron configurations $E_1$ to $E_4$ in a two-orbital model, with their corresponding energies denoted underneath. (b) Theoretical estimates of the intra- and inter-orbital screened Coulomb interactions $W$ and $W'$, as well as their difference $W - W'$, as a function of the substrate dielectric constant $\varepsilon_{\text{bot}}$. (c) Corresponding estimates for the quasi-particle gap $E_g$, particle-hole excitation energy $\Delta_c$, and exciton binding energy $E_{\text{ex}}$.}
\label{fig:excitonModel}
\end{figure}

To theoretically understand why the electronic structure is sensitive to Coulomb engineering, but the CDW formation is not, we consider a simplified model consisting of two orbitals with energies $\varepsilon_c$ and $\varepsilon_v$, representing the conduction band with Ti-$d$ character and the valence band with Se-$p$ character, respectively (Fig.~\ref{fig:excitonModel}(a)). 
The minimal Hamiltonian takes the form
\begin{align}
\label{eq:ed}
 H = 
 &\sum_{\sigma,\alpha} 
 \varepsilon_\alpha n_{\alpha,\sigma}
 +W\sum_{\alpha} n_{\alpha,\up}n_{\alpha,\dn} + W'\sum_{\sigma,\sigma'} n_{c,\sigma}n_{v,\sigma'},
\end{align}
where $\alpha\in \{c,v\}$, $\sigma,\sigma'\in\{\up,\dn\}$, and $n_{\alpha,\sigma}$ denotes the density operator.
We set the single-particle energies of the Ti and Se orbitals to $\varepsilon_c=88$\,meV and $\varepsilon_v=-12$\,meV, respectively.
From the modelling of the quasi-particle band structure we define the intraorbital interaction as $W = \sum_{\mathbf{q}} W_{\text{TiTi}}(\mathbf{q})$, the interorbital interaction as $W' = \sum_{\mathbf{q}} W_{\text{TiSe}}(\mathbf{q})$ and assume all other Coulomb matrix elements to be zero. In Fig.~\ref{fig:excitonModel}(b) we show how $W$ and $W'$, as well as their difference $W-W'$, depend on the substrate dielectric constant $\varepsilon_{\text{bot}}$. Notably, the dipole term $W-W'$ is relatively weakly dependent on $\varepsilon_{\text{bot}}$ as compared to the monopole terms $W$ and $W'$.
In order to estimate the effect of the substrate on CDW formation, we consider the phonon softening described by the following phonon self-energy in the static limit 
\begin{equation}
    \Pi(\mathbf{q},\omega=0) = g^2 \chi(\mathbf{q},\omega=0) \approx \frac{2 g^2}{\Delta_c},
\end{equation}
where $g$ is the electron-phonon coupling strength and $\chi(\mathbf{q},\omega=0)$ is the charge susceptibility \cite{berges2023phonon}. In the second equality we inserted the charge susceptibility evaluated in our simple model at zero temperature, where $\Delta_c$ is the cost of electron-hole excitations and the factor of 2 is from spin degeneracy. From the above expression we understand that the phonon self-energy that is necessary for the CDW formation is inversely proportional to $\Delta_c$, such that in the following we can limit our discussion to the dependence of $\Delta_c$ on environmental screening.

To this end, we closely follow Ref.~\cite{van_loon_RPA_2021} and consider the electron configurations shown in Fig.~\ref{fig:excitonModel}(a). In the ground state $E_1$, the Se orbital is completely filled and the Ti orbital is completely empty. Removing or adding an electron from this state yields energies $E_2$ and $E_3$, respectively. An estimate for the electronic quasiparticle gap $E_g$ as probed in ARPES is then the difference between the electron ionization ($E_2 - E_1$) and affinity ($E_1 - E_3$) energy, such that
\begin{align}
    E_g 
    &= (E_2 - E_1)  - (E_1 - E_3) \notag\\
    &= \varepsilon_c-\varepsilon_v-(W-W')+W'.
\end{align}
The cost of a particle-hole excitation $\Delta_c = E_4 - E_1$ is
\begin{align}
    \Delta_c = \varepsilon_c-\varepsilon_v - (W - W') < E_g,
\end{align}
which is smaller than the quasi-particle gap $E_g$ as a consequence of the attractive interaction between the electron and the hole~\cite{van_loon_RPA_2021}. This yields an estimate for the exciton binding energy given by
\begin{equation}
    E_{\text{ex}} = E_g - \Delta_c = W'.
\end{equation}
We show $E_g$, $\Delta_c$, and $E_{\text{ex}}$ as a function of $\varepsilon_{\text{bot}}$ in Fig.~\ref{fig:excitonModel}(c) [note that the values of $E_g$ are different from those in Fig.~\ref{fig:fig3}(g) due to the simplified model used here]. We find that both the quasi-particle gap $E_g$ and the exciton binding energy $E_{\text{ex}}$ are efficiently tuned by Coulomb engineering following essentially $W'$. Interestingly, we find that this tunability is not reflected by the cost of particle-hole excitations $\Delta_c$, where both $W'$ dependencies cancel out and only the weaker dependence of the dipole term $\sim(W-W')$ enters. As a result, $\Delta_c$ is only weakly dependent on $\varepsilon_{\text{bot}}$. The transition to the CDW phase should therefore be relatively insensitive to Coulomb engineering because the screening effects on $E_g$ and $E_{\text{ex}}$ mostly cancel each other. The relative excitonic contribution to $\Delta_c$ can therefore be tuned by changing the screening. This leaves the intriguing situation where some excitonic contributions can occur in the hBN-supported case, yet yielding the same $T_{\mathrm{CDW}}$ as for the graphite-supported case where we have shown that the phase transition occurs even without an excitonic contribution.

While our study therefore shows that excitonic contributions to the phonon softening can be effectively tuned by Coloumb engineering here, it also directly indicates that the CDW remains stable when these are suppressed, realising an intriguing, but ultimately conventional, charge density wave in TiSe$_2$. This finding is in good agreement with very recent high-fidelity electronic structure calculations for bulk TiSe$_2$, performed using quasiparticle self-consistent GW calculations which were extended to include ladder diagrams in the screened Coulomb interaction~\cite{Larsen2024}. These point to a negligible contribution of electron-hole interactions in the higher screening environment of the bulk system. Moreover, using molecular dynamics calculations, they predicted that lattice fluctuations can produce significant CDW-like signatures even at room temperature, far above the CDW transition \cite{Larsen2024}, and again in excellent agreement with our observation of persistent backfolded spectral weight to temperatures of approximately twice $T_c$ in our experimental measurements.

We note that multiple systems have been proposed in recent years as putative excitonic insulator candidates~\cite{Wakisaka2009,Lu2017,Jia2022,Sun2022, Song2023,Gao2023,Gao2024, Huang2024}, several based on their similarity to the electronic structure of TiSe$_2$~\cite{Song2023,Gao2023,Gao2024}. Our results here suggest that extreme care should be taken when assigning capacity to host the exciton insulator phase based on band structure arguments alone. More broadly, our results directly demonstrate the Coulomb engineering of the electronic structure of TiSe$_2$, allowed by our hybrid exfoliation-epitaxy materials advance. This, in turn, opens the exciting prospect for deterministically modifying long-range Coulomb interactions for tuning the many-body states and phases of a broad array of 2D quantum materials.

\section*{Acknowledgements}
We thank Matt Watson and Peter Wahl for useful discussions. 
We gratefully acknowledge support from the Engineering and Physical Sciences Research Council (under Grant Nos.~EP/X015556/1 and EP/T02108X/1) and the Leverhulme Trust (Grant No.~RL-2016-006).
We gratefully acknowledge MAX IV Laboratory for time on the Bloch beamline under Proposal Nos. 20220959, 20230206 and 20231118. Research conducted at MAX IV, a Swedish national user facility, is supported by the Swedish Research council under contract 2018-07152, the Swedish Governmental Agency for Innovation Systems under contract 2018-04969, and Formas under contract 2019-02496.
Y.V. and T.W. acknowledge funding by the Cluster of Excellence ‘CUI: Advanced Imaging of Matter’ of the Deutsche Forschungsgemeinschaft (DFG) (EXC 2056, Project ID 390715994) and the DFG research unit FOR 5249 (‘QUAST’, Project No. 449872909). J.B. acknowledges funding by 
the DFG under Germany's Excellence Strategy (University Allowance, (EXC
2077, Project ID 390741603, University of Bremen)). For the purpose of open access, the authors have applied a Creative Commons Attribution (CC BY) licence to any Author Accepted Manuscript version arising. 

\section*{Data Availability Statement}
The research data supporting this publication will be made publicly available at [DOI TO BE INSERTED].

\appendix
\section{Methods}
\subsection{Sample prearation}
For the hBN substrates, hBN flakes were mechanically exfoliated from bulk crystals (purchased from \textit{hqgraphene}) using \textit{Nitto} wafer tape, and directly transferred to a 0.5\,wt\% Nb:TiO\textsubscript{2} substrate (purchased from \textit{Shinkosha}). The substrates were then cleaned using acetone and isopropanol before loading them into the UHV system. 
For the graphite substrates, natural graphite single crystals purchased from \textit{NGS Naturgraphit} were used. For easier mounting, the crystals were glued onto tantalum chips using high temperature carbon paste and annealed in a separate vacuum chamber at $700^\circ\mathrm{C}$ and $\sim10^{-8}$\,mbar for several hours to outgas the glue. The graphite crystals were then cleaved in air using adhesive tape and immediately loaded into the ultra-high vacuum (UHV) system for film deposition.
All substrates were annealed in the load lock of the UHV system at $200^\circ\mathrm{C}$ for $\sim$10\,h and again in the MBE chamber at 700--800$^\circ\mathrm{C}$ directly before the growth. The freshly cleaved graphite crystals were annealed for $\sim30$\,min whereas the hBN substrates were annealed for at least $1\,$h and always at $800^\circ\mathrm{C}$ in order to minimise any impurities from the mechanical exfoliation.

The film growth was conducted using the nucleation-assisted growth mode described in Ref.~\cite{Rajan2024}. During the growth, the samples were held at $550^\circ\mathrm{C}$ with a Se beam equivalent pressure of $2\times10^{-7}$\,mbar. After growth, the samples on hBN substrates were annealed under this Se pressure and at growth temperature for up to 1\,h. Finally, they were cooled down from growth temperature at 20\,K/min and the Se flux was shut at $260^\circ\mathrm{C}$. For samples grown on graphite substrates, post growth annealing leads to excessive monolayer to bilayer conversion, and so the samples were directly cooled down under Se flux. For a near complete monolayer, the growth time on hBN was 2\,h~40\,min and on graphite 1\,h~5\,min. The base pressure of the growth chamber was $2\times10^{-10}$\,mbar and typical background pressures during the deposition are 2--4$\times10^{-9}$\,mbar.

After the growth, the samples were capped with a $\sim5$~nm thick layer of amorphous Se, deposited at room temperature. This acted as a protective cap for transporting the samples through air for the ARPES measurements. The cap was removed {\it in situ} in the measurement system by mild annealing at $265^\circ\mathrm{C}$ for 1\,h. This decapping procedure was optimised based on XPS measurements of the Se\,3$d$ core level (see Supplementary Fig. 1). The decapping temperature was monitored with a thermocouple located next to the sample plate, and the pressure during decapping was kept below $1\cdot10^{-9}\,$mbar.

\subsection{ARPES}
ARPES experiments were performed at the Bloch A-branch endstation of the MAX IV synchrotron, Lund, Sweden. A photon energy of 85\,eV and LV polarisation was used in all measurements, which we found  maximised the spectral weight at the top of the valence band, due to photoemission matrix element variations. For Fermi level referencing of the samples grown on hBN, which still exhibited a small amount of charging, we measured the Se\,3$d$ core levels together with each valence spectrum across the temperature series. To avoid changing the beam flux on the sample, these were measured using the second harmonic of the core 85~eV photons. The Fermi level was determined for the base temperature spectrum by fitting of a Fermi function to the data, with the charging-related broadening accounted for with a Gaussian convolution. The Fermi level for the higher temperature spectra was then calculated based on the shift of the Se core levels across the temperature series. To correct for thermal drift of the samples, we performed spatial re-registration using ARPES-based spatial mapping. We found that the degree of twin domain formation showed some spatial variations on the sample, leading to a change in relative intensity at positive and negative momenta. To rule out any influence of this for our spectral weight analysis, we thus performed the ROI analysis by taking the average of the integrated intensity of the backfolded weight at positive and negative momenta. We then subtracted the background intensity and normalised the intensity to the background-corrected value to obtain the replica intensity reported in Fig.~\ref{fig:fig4}(e). For the estimation of the free-carrier density in TiSe\textsubscript{2}/hBN, we assume a nearly-free-electron-like dispersion of the conduction band with an effective mass based on the TiSe\textsubscript{2}/graphite data.

\subsection{Quasi-particle band gap}
For the theoretical modelling of the Coulomb engineering of the band gap, we start from the Wannierized density functional theory (DFT) bandstructure of ML-TiSe$_2$ in Ref.~\cite{schobert2024ab}. It is constructed from 5 Ti $3d$-orbitals and $2\times 3$ Se $4p$-orbitals, which together dominantly contribute to the bands around the Fermi energy. 
DFT predicts ML-TiSe$_2$ to be metallic due to the well-known bandgap problem of semi-local DFT functionals. To correct for this, we use the scissor operator to introduce a gap of $0.1$\,eV (i.e., by shifting the Ti $d$- and Se $p$-orbitals up and down in energy, respectively), leading to the bandstructure shown in Supplementary Fig.~3.

To analyse the effect of Coulomb engineering, we augment this model with Coulomb interactions.
The bare Coulomb interaction in the density-density approximation is written as
\begin{equation}
    V_{\alpha\beta}(\mathbf{q}) = \frac{2 \pi e^2}{A q} e^{-i \mathbf{q} \cdot \mathbf{\tau}_{\alpha\beta}} e^{- q d_{\alpha\beta}},
\end{equation}
where $\mathbf{\tau}_{\alpha\beta}$ is the in-plane distance between orbitals $\alpha$ and $\beta$, and $d_{\alpha\beta}$ the corresponding out-of-plane distance. For simplicity, we set the orbital positions to the positions of the corresponding atoms.
Similar to Ref.~\cite{rosner2015wannier}, we treat screening from the environment, as well as internal screening from bands outside of the low-energy model, by dividing the leading eigenvalue of $V_{\alpha\beta}(\mathbf{q})$ by
\begin{equation}
    \varepsilon_{\text{back}}(\mathbf{q}) = \varepsilon_{\text{int}} \frac
    {1 - \beta_{\text{top}} \beta_{\text{bot}} e^{-2 q h}}
    {1 + (\beta_{\text{top}} + \beta_{\text{bot}}) e^{-q h} + \beta_{\text{top}} \beta_{\text{bot}} e^{-2 q h}},
\end{equation}
where $\beta_{\text{top/bot}} = (\varepsilon_{\text{int}} - \varepsilon_{\text{top/bot}})/(\varepsilon_{\text{int}} + \varepsilon_{\text{top/bot}})$, and dividing the subleading eigenvalues by $\varepsilon_{\text{int}}$. The resulting background screened Coulomb interaction is denoted by $U_{\alpha\beta}(\mathbf{q})$. Here we set the internal TiSe$_2$ dielectric constant $\varepsilon_{\text{int}} = 60$. The dielectric constant of the substrate $\varepsilon_{\text{bot}}$
 is set to $3$ for TiSe$_2$ mounted on hBN, or to infinity when mounted on graphite. We set the dielectric constant above the TiSe$_2$ layer to $\varepsilon_{\text{top}} = 1$, representing vacuum, and the effective TiSe$_2$ thickness to $h = 8$\,\AA.

To qualitatively describe the renormalization of the normal-state due to non-local Coulomb interactions, we use the Coulomb hole plus screened exchange (COHSX) approximation~\cite{Hedin1965,Steinke2020}. It is given by the following self-energy in the orbital basis
\begin{widetext}
\begin{align}
    \Sigma^{\text{COHSX}}_{\alpha\beta}(\mathbf{k}) =&
    -\sum_{\mathbf{k}',i} T_{\alpha i}(\mathbf{k}') T_{i\beta}^{\dagger}(\mathbf{k}') n_F(\xi_{\mathbf{k}',i}) W_{\alpha\beta}(\mathbf{k}-\mathbf{k}',\omega=0) + \frac{1}{2} \delta_{\alpha\beta} \sum_{\mathbf{q}} \left( W_{\alpha\alpha}(\mathbf{q},\omega=0) - U_{\alpha\alpha}(\mathbf{q}) \right),
\end{align}
\end{widetext}
where $T_{\alpha i}(\mathbf{k})$ are the unitary transformation matrices that transform the bare electron Hamiltonian to the band basis $\xi_{\mathbf{k},i}$, and $n_F(\varepsilon)$ is the Fermi-Dirac distribution function. The statically screened Coulomb interaction in the density-density approximation $W_{\alpha\beta}(\mathbf{q},\omega=0)$ is obtained from the random phase approximation (RPA). The RPA and COHSX calculations have been performed using the TRIQS~\cite{parcollet2015triqs} and TPRF~\cite{strand2023tprf} codebases, with linearly discretized momentum meshes with $100\times100$ points.

We show the quasi-particle bandstructure in the COHSX approximation in Supplementary Fig.~3 for $\varepsilon_{\text{bot}} = 3$ and $\varepsilon_{\text{bot}} = \infty$, representing hBN and graphite substrates, respectively. We find a rigid upward shift of the conduction band, accompanied by a rigid downward shift of the valence band, due to the effects of the self-energy $\Sigma^{\text{COHSX}}$.
The resulting bandgap enhancement is larger for the hBN substrate than for the graphite substrate as a consequence of the reduced screening, as also shown in Fig.~\ref{fig:fig3}(g).

\

\foreach \x in {1,...,4}
{%
\begin{figure*}[!ht]
  \centering
  \includegraphics[
  width=\textwidth,
  page=\x,
  trim=1.5cm 1cm 1.5cm 1.5cm,
  clip
]{supp.pdf}
\end{figure*}
}

\end{document}